\documentstyle[12pt]{article}

\catcode`\*=11  

\def\sl*sh#1#2#3{\ooalign{\setbox0=\hbox{$#2\not$}
                          $\hfil#2\mkern-24mu\mkern#1mu
                           \raise.15\ht0\box0\hfil$\cr
                          $#2#3$}}
\catcode`\*=12
\newcommand{\beq}{\begin{equation}}
\newcommand{\eeq}{\end{equation}}
\newcommand{\beqa}{\begin{eqnarray}}
\newcommand{\eeqa}{\end{eqnarray}}

\newcommand{\eq}[1]{(\ref{#1})}

\newcommand{\ra}{\rightarrow}
\newcommand{\del}{\partial}
\newcommand{\tr}{{\rm Tr}}
\newcommand{\sh}{{\rm sh}}

\newcommand{\teta}{\Theta_2(s{2\mu\over\beta},is{4\pi\over\beta^2})}
\newcommand{\integ}{\int_0^\infty}

\newcommand{\NP}[1]{ {\it Nucl.~Phys.} {\bf #1}}
\newcommand{\PL}[1]{ {\it Phys.~Lett.} {\bf #1}}
\newcommand{\Prep}[1]{ {\it Phys.~Rep.} {\bf #1}}
\newcommand{\PR}[1]{ {\it Phys.~Rev.} {\bf #1}}

\newcommand{\MPL}[1]{ {\it Mod.~Phys.~Lett.} {\bf #1}}
\newcommand{\IJMP}[1]{ {\it Int.~Jour.~Mod.~Phys.} {\bf #1}}

\newcommand{\TMP}[1]{ {\it Theor.~Math.~Phys.} {\bf #1}}
\begin{document}
\topmargin 0pt
\oddsidemargin 1mm
\begin{titlepage}

\rightline{  }
\setcounter{page}{0}
\vspace{15mm}
\begin{center}
{\Large  On thermal phase structure of deformed Gross-Neveu model} 
\vspace{20mm}

{\large H-T. Sato$^{b)}$  
\footnote{ Fellow of the Danish Research Academy,{ }~ 
 sato@nbi.dk}} 
{\sc and }
{\large H. Tochimura$^{a)}$
\footnote{tochi@theo.phys.sci.hiroshima-u.ac.jp} }\\
\vspace{10mm}

$^{a)}${\em Department of Physics, Hiroshima University\\
        Kagamiyama 1-3-1, Higashi-Hiroshima 739, Japan}\\
$^{b)}${\em The Niels Bohr Institute, University of Copenhagen\\
     Blegdamsvej 17, DK-2100 Copenhagen, Denmark}\\
\end{center}
\vspace{7mm}

\begin{abstract}
We illustrate the phase structure of a deformed two-dimensional 
Gross-Neveu model which is defined by undeformed field contents 
plus deformed Pauli matrices. This deformation is based on two 
motives to find a more general polymer model and to estimate 
how $q$-deformed field theory affects on its effective potential. 
There found some regions where chiral symmetry breaking and 
restoration take place repeatedly as temperature increasing. 
\end{abstract}

\vspace{1cm}

\end{titlepage}
\newpage
\renewcommand{\thefootnote}{\arabic{footnote}}
\indent

Gross-Neveu (GN) model and its relevant models have been studied 
intensively as a simple, but clear, model of chiral symmetry breaking 
in various situations; under the circumstances of thermal \cite{therm}, 
electromagnetic \cite{HYP,gauge} or curvature \cite{curve} backgrounds 
(and references are therein). 

Another plausible application of these (two-dimensional) models is 
an effective theory to one-dimensional systems such as a polymer 
(for a review see \cite{rev}). It is shown in \cite{poly} that the 
$N=2$ GN model is equivalent to a continuum model of polyacetylene. 
{}~Further, the first order phase transition of polyacetylene 
from solitonic (chiral broken) phase to metallic (symmetric) phase of 
is formulated as a response to the doping concentration (chemical 
potential), {\it i.e.} the number of dopants per carbon atom, 
using large $N$ approximation \cite{CM}. The critical value of 
chemical potential coincides with an experimental value (Recently, 
the $1/N$ correction to this model has been calculated \cite{CM2}). 
There is also a discretized version (fermionic vector model 
\cite{vector}) which has been studied as one of polymer models on 
discretized surfaces (random polymers) or as string theory and lower 
dimensional quantum gravity \cite{matrix}. 

 As shall be shown later, we immediately see that the effective 
potential of the two-dimensional GN model (defined in the large $N$ 
leading order) under a static background corresponds to the electrostatic 
potential of a specific {\it charge} distribution on a half-line. 
Consequently, solitonic-metallic phase transitions of a polymer 
can be regarded as phase transitions concerning its electrostatic 
states as well. This suggests that there exist polymer models as many 
as electrostatic potentials.
On the one hand, one may expect a universal structure of electric 
charge distributions on a polymer. On the other hand, we have only a 
few GN models coupled to background fields, which play the role of 
{\it charge} distributions. 
Hence, we introduce a new example of electrostatic potential, which 
is under an oscillatory distribution, deforming the effective potential 
of the GN model coupled to a constant background gauge field. 
This approach would broaden the range of application of model within the 
method of large $N$ calculation. Since we deform the Pauli matrices, 
our deformation may correspond to a kind of fluctuating noise, 
quon particles \cite{quon} or a violation effect similar to the Pauli 
principle violation \cite{pauli}, which 
are formulated using $q$-deformation. 

{}~From a viewpoint of high energy physics, the deformation possesses 
the following meaning. Motivated by unusual spacetime structure like 
non-commutative geometry \cite{noncom}, which would be anticipated in 
an extremely high energy scale, there are several $q$-deformed field 
theory constructions \cite{qfield}. We would like to focus on a pure 
effect from $q$-deformation rather than from gravitational effect here. 
However, all of these deformed models are not yet viable to calculate 
effective actions performing path integrals because of their 
mathematical intricacy --- fermions belong to quantum Lorentz group, 
and deformed Pauli matrices are defined by multiplications of 
$q$-factors which are governed by an appropriate quantum $R$-matrix 
\cite{qgamma}. 
In order to catch a glimpse of $q$-deformed field theory aspect, we 
employ a likely approximation which definitely makes calculation 
possible. Namely, we will use ordinary (undeformed) fermions plus 
a common $q$-factor ($q=e^{\epsilon\pi i}$) th the Pauli matrices. 

In this report, we investigate the thermodynamics and its phase 
structure of the deformed GN model. If we take the vanishing limit 
of an electric background (or of the deformation parameter), our 
model is exactly reduced to the thermal generalization of the 
original model discussed in \cite{CM}. The phase diagrams we shall 
show here cover every known phase boundary of the undeformed GN 
model accordingly. One of our interests is thereby how the only 1st 
order transition point extends into a finite region of phase space.

Our starting model Lagrangian (before deforming) is 
\beq
{\cal L}={\bar\psi}i\gamma^\mu(\partial_\mu-ieA_\mu)\psi
-{N\over2\lambda}\sigma^2-\sigma{\bar\psi}\psi, \label{basemodel}
\eeq
where $N$ stands for the number of flavors, and summations 
over flavors are implicit. $A_\mu$ is a constant background field.  
The large $N$ effective (bare, $D=2$) potential of this model is 
expressed in the following form
\beq
V(\sigma;\xi)-V(0,\xi) ={1\over2\lambda}\sigma^2+
            \integ ds {1\over 4\pi s^2} Q(s\xi) e^{-s\sigma^2},
\label{no1}
\eeq
where $\xi$ is a background constant as we shall see later. 
The second term (quantum correction) on r.h.s. of \eq{no1} can be 
interpreted as the electrostatic potential of an effective charge 
(form factor) $Q$ with damping factor $e^{-s\sigma^2}$ on the  
half-line parametrized by $s$, 
while the first (classical) term is a constant in a sense. Thus, 
the function $Qe^{-s\sigma^2}$ represents a charge distribution in 
one dimension, and $\sigma^{-2}$ corresponds to an effective range 
of electric force. In the metallic (chiral symmetric) phase, 
$<\sigma>=0$, the effective range becomes infinity, and the 
potential behaves as the Coulomb potential of charge $Q$. $\xi$ 
plays the role of {\it einbein} as it stands in \eq{no1}, but we can 
see its meaning more clearly in other models. For example, in the 
higher dimensional (magnetic dominant) cases of \eq{basemodel}, 
$Q$ and $\xi$ are given by \cite{gauge}
\beq 
  Q(s)= s {\rm coth}(s), \quad \xi= e\sqrt{-E^2 +B^2}, \label{no2}
\eeq
where $E$ is assumed to be small and perpendicular to $B$. When $E=0$, 
$\xi^{-1}$ is called the magnetic length which represents the radius of 
cyclotron motion of each electron in classical mechanical picture, and 
hence electrons are spaced by $\xi^{-1}$. 
In the case of constant curvature background, we have, 
for example \cite{KS},  
\beq
Q(s)=e^{-s/2} 
             \left( { s/D\over \sh(s/D)} \right)^{1/2}
  \left( {s/D(D-1)\over \sh(s/D(D-1))} \right)^{D-1\over2}, \quad
\xi={R\over2},                                      \label{no3}
\eeq  
where a weak curvature approximation is assumed.
In each case, $\xi^{-1}$ represents a specific length of the system 
and plays the role of {\it einbein} in the proper-time integral. 
Owing to this property, we may suppose a lattice system of spacing 
$\xi^{-1}$, and this fact might have a relation to carbon lattice 
systems.

Let us go back to the $D=2$ story. Obviously, in chiral symmetry 
breaking theories, there is no sense of taking $D=2$ limit in 
\eq{no2} because of no magnetic field in two dimensions. However, 
as mentioned previously, it still makes sense as a problem of 
1-dimensional electrostatics and generalization of polymer model, 
since what functional form of $Q$ we choose corresponds to a 
definition (deformation) of polymer model. {}~Furthermore, from a 
field theoretical viewpoint, $D=2$ is just a toy model which 
simplifies most of equations, however dimensionality is not a 
crucial factor for the purpose of observing a pure effect of 
$q$-deformation. One can repeat the same analysis in higher 
dimensions. We therefore define every quantity of $D=2$ theories 
in the formal limit $D\ra2$.

In order to study an oscillatory function, we further deform 
\eq{no2}, and our choice of $Q$ is 
\beq
       Q(s) = s {{\rm cos } s \over {\rm sh } s}.   \label{no7}
\eeq 
Let us briefly see why the choice is a deformation of \eq{no2}. 
In the process of deriving \eq{no2}, using the dimensional 
regularization, we have introduced the following slight phase 
difference between the gamma matrices and the auxiliary field $\sigma$
\beq 
   \sigma\quad\ra\quad\sigma, \hskip 20pt
 \gamma_{\mu} \quad\ra\quad e^{\epsilon\pi i} \gamma_{\mu}. \label{no4}
\eeq
This procedure brings a slight violation of chiral symmetry (and of 
course the replacements of $\gamma_\mu$ in \eq{basemodel}). This gives 
a naive $q$-deformation ($q=e^{\epsilon\pi i}$) of the gamma matrices 
in the $D=2$ model. 
Since $Q$ of eq.\eq{no2} is originally proportional to \cite{HYP}
\beq
\tr\exp[ {i\over2}esF_{\mu\nu}\sigma^{\mu\nu} ]
\times \exp[-\tr\left(\ln{{\rm{sh}}(esF)\over esF}\right)_{\mu\nu}], 
\eeq
the phase deformation \eq{no4} is relevant to only the first trace 
(over the gamma matrices). Here we extract particular two cases 
concerning the first trace, 
\beq
\left\{ 
\begin{array}{ll}
 \mbox{cosh}(es\xi) &\quad\mbox{when} \quad \epsilon=n/2,  
  \qquad(n\in Z)\\
 \mbox{cos}(es\xi) &\quad \mbox{when} \quad \epsilon=(2n+1)/4. 
\end{array}\right.
\eeq
Nothing happens to the charge distribution function in the former case, 
while the latter case gives rise to an oscillatory function without 
any singularity. When $\epsilon$ is an integer, which corresponds to 
the undeformed model, we reproduce the theory \eq{no2}. 
This case will be reported in a forthcoming paper. 
Choosing the latter case as a deformed GN model, we arrive at \eq{no7}. 
We should consider $\xi$ as a specific electric length under the charge 
distribution given by \eq{no7}.

Now, we are in a position to discuss the thermodynamic potential of 
our model defined by \eq{no1} with eq.\eq{no7} which is deformed from 
the theory \eq{no2} and is defined in the formal limit $D\ra2$ as 
already mentioned. The thermodynamic potential is obtained in the 
following form applying the method of \cite{KS} to our model
(notation and derivation are completely parallel to \cite{KS})
\beq
V(\sigma;\xi,\beta,\mu)={1\over2\lambda}\sigma^2+
           { {\rm tr}[1] \over2\beta}
            \integ {ds\over s}{\teta\over(4\pi s)^{(D-1)/2}}
           Q(s\xi) e^{s\mu^2} (e^{-s\sigma^2}-1),    \label{no8}
\eeq
where $T=\beta^{-1}$, and we have introduced the dimensional 
regularization and $tr[1]$ means the trace of gamma matrix unit. 
The polymer model of \cite{CM} is a particular case of this potential 
($\xi=0, \beta\ra\infty$ and of course $D=2$), 
and the case of $\xi=0$ with a finite $\beta$ is the thermodynamic 
generalization of the polymer model. Note also that the replacements 
of $Q$ with \eq{no2} and \eq{no3} reproduce the thermodynamic 
potentials discussed in \cite{potential} and \cite{KS} respectively. 
Imposing the condition 
\beq
\lim_{T,\mu\ra0} {\del^2\over\del\sigma^2}
V(\sigma;\xi,T,\mu)\Bigr\vert_{\sigma=1} ={1\over\lambda_R}, 
\label{no9}
\eeq
the renormalization of coupling constant $\lambda_R$ is given by 
\beq
{1\over\lambda}-{1\over\lambda_R}={\rm tr}[1] \integ {ds\over
       (4\pi s)^{D/2}}e^{-s}(1-2s)Q(s\xi),               \label{no10}
\eeq
and the renormalized effective potential is therefore 
\[
\hskip -55pt
V(\sigma;R,\beta,\mu)={1\over2\lambda_R}\sigma^2
+{1\over2}{\rm tr}[1]\integ {ds\over(4\pi s)^{D/2}} Q(s\xi)
\]
\beq
\hskip 35pt
\times[\,{1\over s}(e^{-s\sigma^2}-1){\sqrt{4\pi s}\over\beta}
e^{s\mu^2}\teta+\sigma^2e^{-s}(1-2s)\,].                   \label{no11}
\eeq
In principle, phase transitions are classified by behaviour of 
potential minima, i.e., by the gap equation $\del V/\del\sigma=0$,
\beq
0={1\over\lambda_R}+{\rm tr}[1]\integ {ds\over(4\pi s)^{D/2}} 
   Q(s\xi)
[\,-e^{-s(\sigma^2-\mu^2)}{\sqrt{4\pi s}\over\beta}\teta + 
                        e^{-s}(1-2s)\,].                   \label{no12}
\eeq
(We hereafter adopt the following value of the renormalized coupling 
constant
\beq
{1\over\lambda_R}=
      {\rm tr}[1]\integ {ds\over(4\pi s)^{D/2}}2se^{-s}, \label{no13}
\eeq
which means the broken phase with the dynamical mass $\sigma=1$ at 
$T=\mu=\xi=0$.) 

However, there are more convenient equations which can be 
derived from the gap equation. These are in order: \par
\noindent 
(i) {\it 2nd order critical surface} satisfies
\beq
0=\integ{ds\over(4\pi s)^{D/2}} [\, Q(s\xi)
  \{ -e^{s\mu^2}{\sqrt{4\pi s}\over\beta}\teta 
  +  e^{-s}(1-2s) \}  + 2se^{-s} \, ],                     \label{no14}
\eeq 
defined by
\beq
\lim_{\sigma\ra0} {\del\over\del\sigma^2}
                 V(\sigma;\xi,\beta,\mu) =0.      
\eeq 
(ii) {\it 3rd order critical line} 
is determined by simultaneous solution of \eq{no14} and 
\beq
    0= \integ{ds\over(4\pi s)^{D/2}} se^{s\mu^2}Q(s\xi)
         {\sqrt{4\pi s}\over\beta}\teta,                  \label{no15}
\eeq
which follows from
\beq
\lim_{\sigma\ra0}({\del\over\del\sigma^2})^2 
                 V(\sigma;\xi,\beta,\mu) =0.               \label{no18}
\eeq
(iii) {\it 4-th critical point} is defined by simultaneous solution of 
\eq{no14}, \eq{no15} and 
\beq
\lim_{\sigma\ra0}({\del\over\del\sigma^2})^3 
                 V(\sigma;\xi,\beta,\mu) =0.               \label{no18}
\eeq
This is alternatively written in the following, but it has no 
solution in our case:
\beq
    0= \integ{ds\over(4\pi s)^{D/2}} s^2e^{s\mu^2}Q(s\xi)
         {\sqrt{4\pi s}\over\beta}\teta.                  \label{no17}
\eeq
Remember that every solution of \eq{no14} is not always on the true 
critical surface. 3rd order critical line cuts off irrelevant solutions 
of \eq{no14} away from the critical surface. Instead, 1st order 
critical surface should be found from simultaneous solution of $V=0$ 
and $\partial V/\partial\sigma=0$. 

Note that eqs.\eq{no14} and \eq{no15} are reduced to the following 
simple equations in some regions: \par
\noindent 
(a) {\it Limit of $\xi\ra0$ } ($T$-$\mu$ plane):
\beq
\beta^{D-2}\Gamma(1-{D\over2})={2\over\sqrt{\pi}} (2\pi)^{D-2}
   \Gamma({3-D\over2}){\rm Re}
   \zeta(3-D,{1\over2}+i{\beta\mu\over2\pi})  \qquad 
          \mbox{for \eq{no14},}      \label{no19}
\eeq
and
\beq
  {\rm Re} \zeta(5-D,{1\over2}+i{\beta\mu\over2\pi})=0  
      \qquad\mbox{for \eq{no15},}   \label{no20}
\eeq
where $\zeta$ is the generalized zeta function. These equations are 
exactly the same as those derived in \cite{Muta}  
where various critical equations \cite{therm},\cite{CM} 
are reproduced from these equations. \par
\noindent
(b) {\it limit of $T\ra0$ } ($\xi$-axis, $\mu=0$):
\beq  
    2\Gamma({D\over2}) \xi^{1-{D\over2}} + 
        \psi_D(\xi^{-1})+{2\over \xi}\psi_D'(\xi^{-1})=0
            \qquad \mbox{for \eq{no14},}  \label{no21}
\eeq
\beq
\xi^{{D\over2}-1}\lim_{z\ra0}{\psi_D(z\xi^{-1})\over z}=0
\qquad \mbox{for \eq{no15},}  \label{no22} 
\eeq
where $\psi_D$ is   
\beq
     \psi_D(z)=\integ ds s^{-D/2} (e^{-sz}-1) Q(s).   \label{no23}
\eeq
\noindent
(c) {\it Leading order in weak $\mu$ expansion}:
\beq
0= \integ{ds\over(4\pi s)^{D/2}} [\, Q(s\xi)
  \{ e^{-s}(1-2s) - e^{s \mu^2}(1+2s\mu^2)^{-1/2} \}
                                   + 2se^{-s} \,]  
\qquad \mbox{for \eq{no14}},    \label{no24}
\eeq
\beq 
    0=\integ {ds\over(4\pi s)^{D/2}}s 
            e^{s \mu^2}(1+2s\mu^2)^{-1/2} Q(s\xi) 
\qquad \mbox{for \eq{no15}} .  \label{no25}
\eeq

The results on our phase structure determined by these equations 
in $D=2$ are summarized in Figs.1-6. In Figs.1-5, we show various 
sections ($\xi$-$T$ planes) of the critical surface from 
$\mu=0$ to 0.7. In Fig.6, we show the solutions of \eq{no24} and 
\eq{no25}. ${\bf S}$ and ${\bf M}$ express solitonic/metallic 
(in the original words, broken/symmetric) phases. On the 
solid/dashed lines, 2nd/1st order phase transitions take place. 
The points $A$, $A_1$, $A_2$, $B$, $B_1$, $B_2$ are the tri-critical 
points which, strictly speaking, form two lines as explained below. 

Let us go into details of how the critical surface looks like in 
order of values of $\mu$.  For the interval $0\leq\mu<0.59$, all the 
phase transitions are second order transitions which satisfy 
\eq{no14}. In Fig.1, we show a representative diagram ($\mu=0$) for 
this interval of $\mu$. Its critical values of $\xi$ and $T$ are 
$\xi_c=0.75$ and $T_c=0.57$. Similar diagrams as Fig.1 appear, 
as $\mu$ increasing whereas $T_c$ decreasing.
When $\mu=0.59$ (Fig.2), we observe the point $A$ which is just a 
turning point of the tricritical line $A_1$-$A$-$A_2$ (see Figs. 2-5). 
This point is not 4-th critical because \eq{no17} has no solution as 
mentioned previously. The coordinates $(\xi,T)$ of $A$ are 
$(0.72,0.08)$. All the transitions of Fig.2 except for the point $A$ 
are second order transitions. 
When $\mu=0.601$ (Fig.3), another tricritical point $B=(0.23,0.3)$ 
appears just similarly as the point $A$. The point $A$ now splits 
into $A_1$ and $A_2$. The dashed line $A_1$-$A_2$ of 1st order 
transitions are determined by solving the equation 
\eq{no11}$=0$ with \eq{no13}. 
In Fig.4 ($\mu=0.607$), we figure out two tricritical lines 
$A_1$-$A$-$A_2$ and $B_1$-$B$-$B_2$ (The dashed lines $A_1$-$A_2$ 
and $B_1$-$B_2$ are of course 1st order). $B_1,B_2,A_1,A_2$ are 
$(0.1,0.31),(0.31,0.285),(0.74,0.094),(0.71,0.07)$ respectively. 
$B_1$ disappears into the negative region of $\xi$ at $\mu=0.608$ 
and $T=0.32$. This is the well-known tricritical point \cite{therm}.  
Finally, in Fig.5, the coordinates of $B_2,A_1,A_2$ of 
$\mu=0.7$ are 
$(0.55,0.24),(0.8,0.108),(0.69,0.059)$ respectively. 

We show in Fig.6 a $\mu$-$\xi$ plane diagram as well, solving \eq{no24} 
and \eq{no25}. In this case, we find a tricritical point 
at $(\mu,\xi)=(0.607,0.695)$ and 1st order transitions in $\mu>0.607$. 
This might mean that the approximation becomes invalid around this point 
(numerical analysis does not work well for some higher value of $\mu$). 
This is a different feature from \cite{KS} which shows only second 
order transitions. 

In conclusion, let us summarize our study. We have 
found, taking a particular charge function, the 1st order transitions 
in certain regions of temperature and the electric length parameter 
$\xi$, whose small value region is continuously connected to the 
known transition point shown in \cite{CM}. This smooth connection of 
1st order transitions would seem to be natural. On the other hand, 
appearance of the tricritical line $A_1$-$A$-$A_2$ (and the concave 
structure of $A_1$-$A_2$) is a main 
difference between our model and the other models \eq{no2} and 
\eq{no3} \cite{KS}. This difference comes out of oscillatory nature 
of the $Q$ function. Although we have observed respective $\xi$-$T$ 
sections of the critical surface, we could have also shown $T$-$\mu$ 
sections at several $\xi$ values. According to Fig.6, assuming 
the first order line to be connected to the origin of Fig.5, 
the critical value of $\mu$ decreases as $\xi$ increasing. This 
feature can be a replacement of other models. 
 
{}~For example, the large $N$ $\xi=0$ model gives a perfect agreement 
with an experimental data and the $1/N$ correction reduces the critical 
chemical potential by $20\%$, which is still acceptable within 
experimental limits \cite{CM2}. Our results of critical chemical 
potentials for $0\leq\xi\leq0.695$ stay within this $20\%$ reduction 
and therefore experimentally acceptable. We can offer various models of 
polyacetylene in conformity with the choice of $\xi$. 

The concave structure of $A_1$-$A_2$ curves shows that the phases 
around there sensitively respond to temperature, and phase transitions 
(mass generations) repeat several times as $T$ increasing. This 
feature can not be seen in the undeformed cases \eq{no2} and \eq{no3}, 
and hence this must be the pure effect of $q$-deformation. If we could 
analyze a more strictly $q$-deformed model, we would observe similar 
unstable phase transitions in some regions of the phase space and in 
more complicated way. Therefore $q$-deformed field theory, as a 
formulation of non-archimedian, non-commutative or foam-like structures  
of spacetime \cite{padic},\cite{foam}, could be related to a certain 
drastically changing phase structure. Although we only studied the 
$D=2$ model, this picture will probably be same even if in 
higher $D$ cases. 

It is still unknown how phase structures 
depend on details of $Q$'s. To know this, there would be no other way 
to collect many examples. Choosing various functional form of $Q$,
it might be interesting to compare how each critical surface differs.  
For example, for the undeformed model, which has another interest of 
chiral symmetry breaking, $Q(s)$ is given by $s{\rm cot}s$ in the 
electric dominant case. In this case, we have to deal with an 
infinite number of singularities of $Q$. Apart from the singularity 
problem, various features of our oscillatory $Q$ might call us a 
precaution when studying phase structures of other oscillatory 
charge functions as well as that singular case.

Finally, we hope that our approximated analyses would be helpful for 
quantitative analyses of strictly $q$-deformed (quantum group 
based) models in both contexts. 


\noindent
{\em Acknowledgments:}
We thank S. Mukaigawa for comment and information.

\vspace{1cm}
\newpage

%
 \pagebreak [4]
 \topmargin 0pt
 \oddsidemargin 5mm
 \centerline{\bf Figure Captions}
 \begin{description}
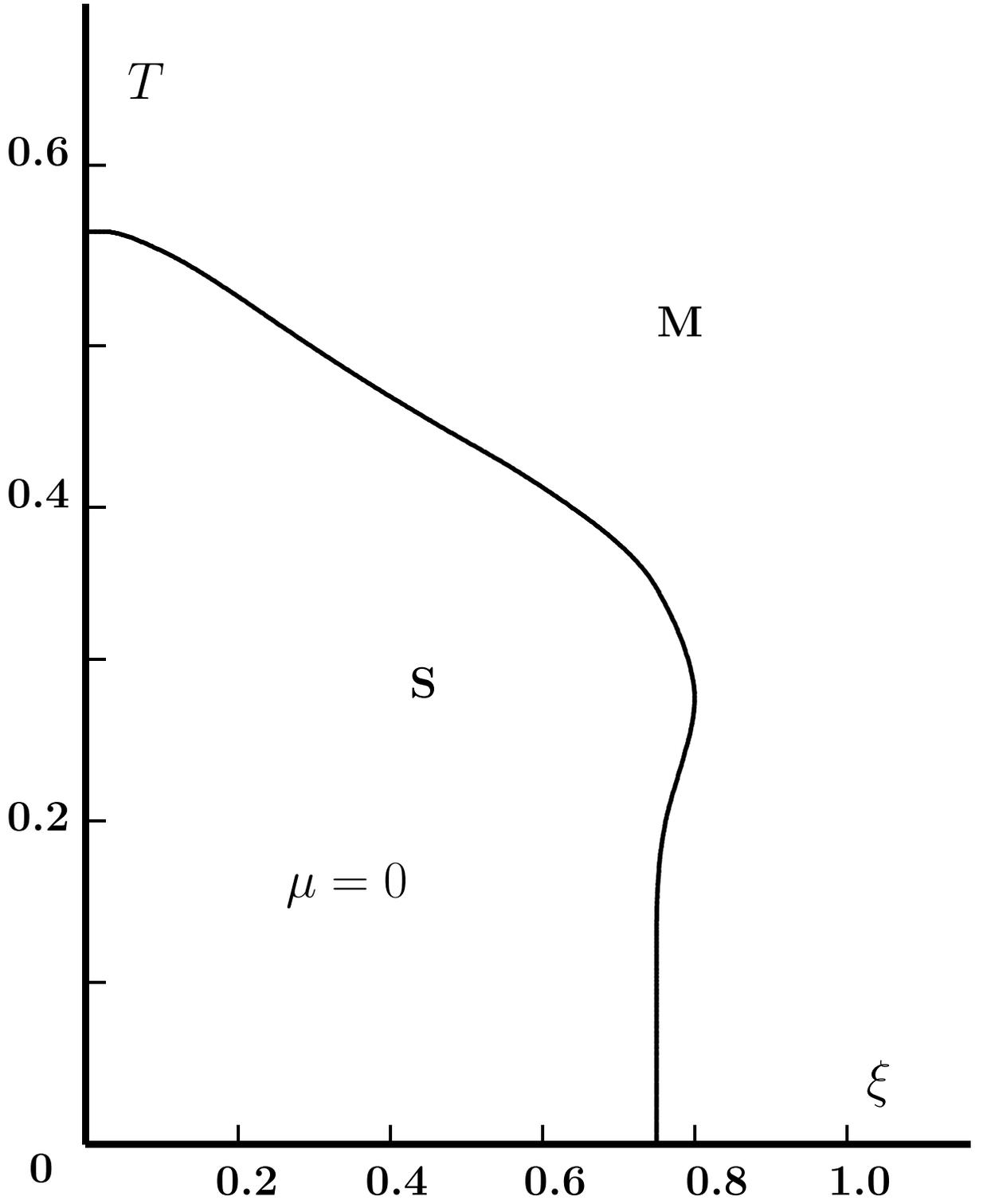
 \item[{\bf Fig.1}:]
 The 2nd order critical line ($\mu=0$) on $\xi$-$T$ plane. 
 ${\bf S}$ and ${\bf M}$ mean solitonic (broken) and metallic (symmetric) 
     phases respectively.
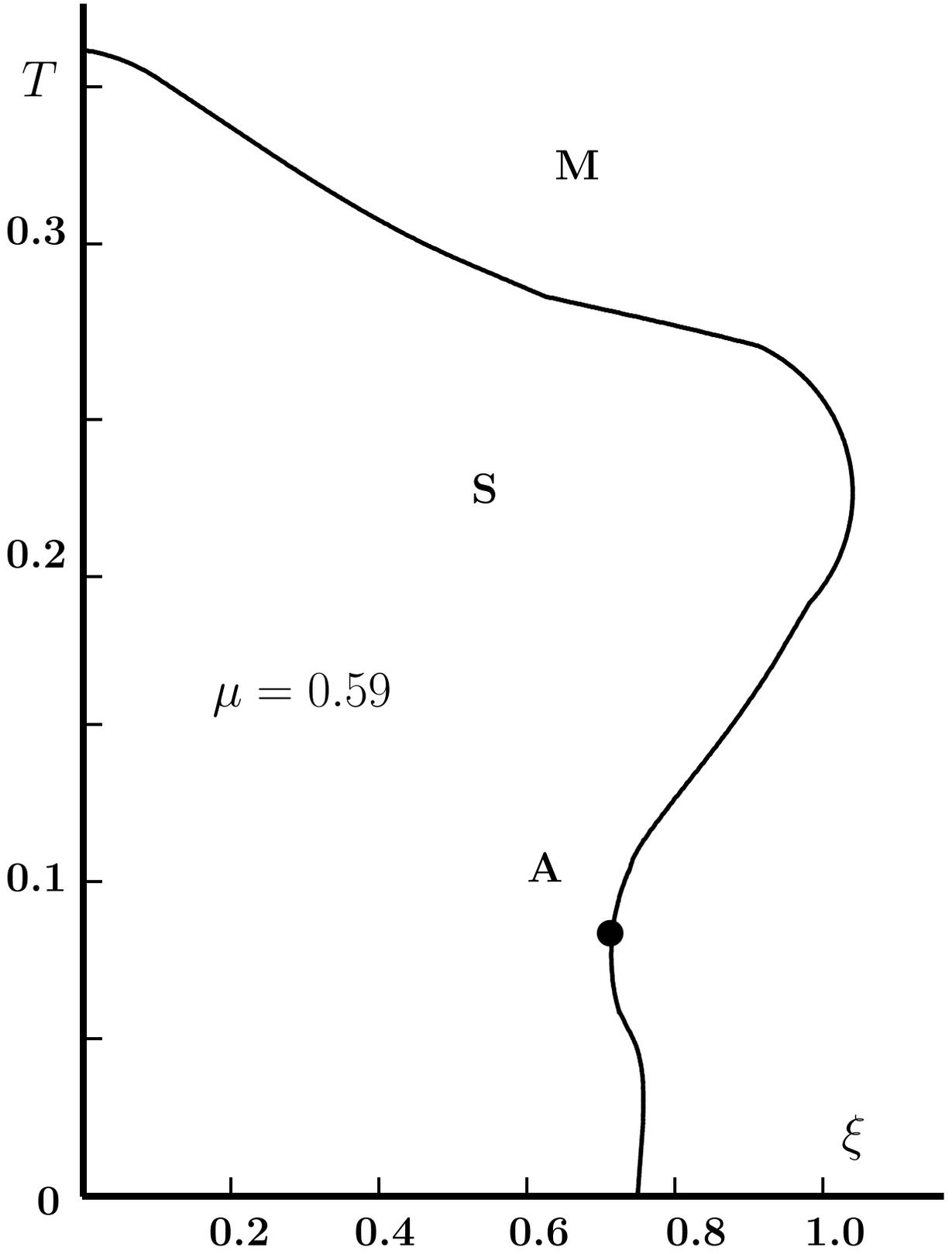
 \item[{\bf Fig.2}:]
 The situation of $\mu=0.59$. ${\bf A}$ is a tricritical point.
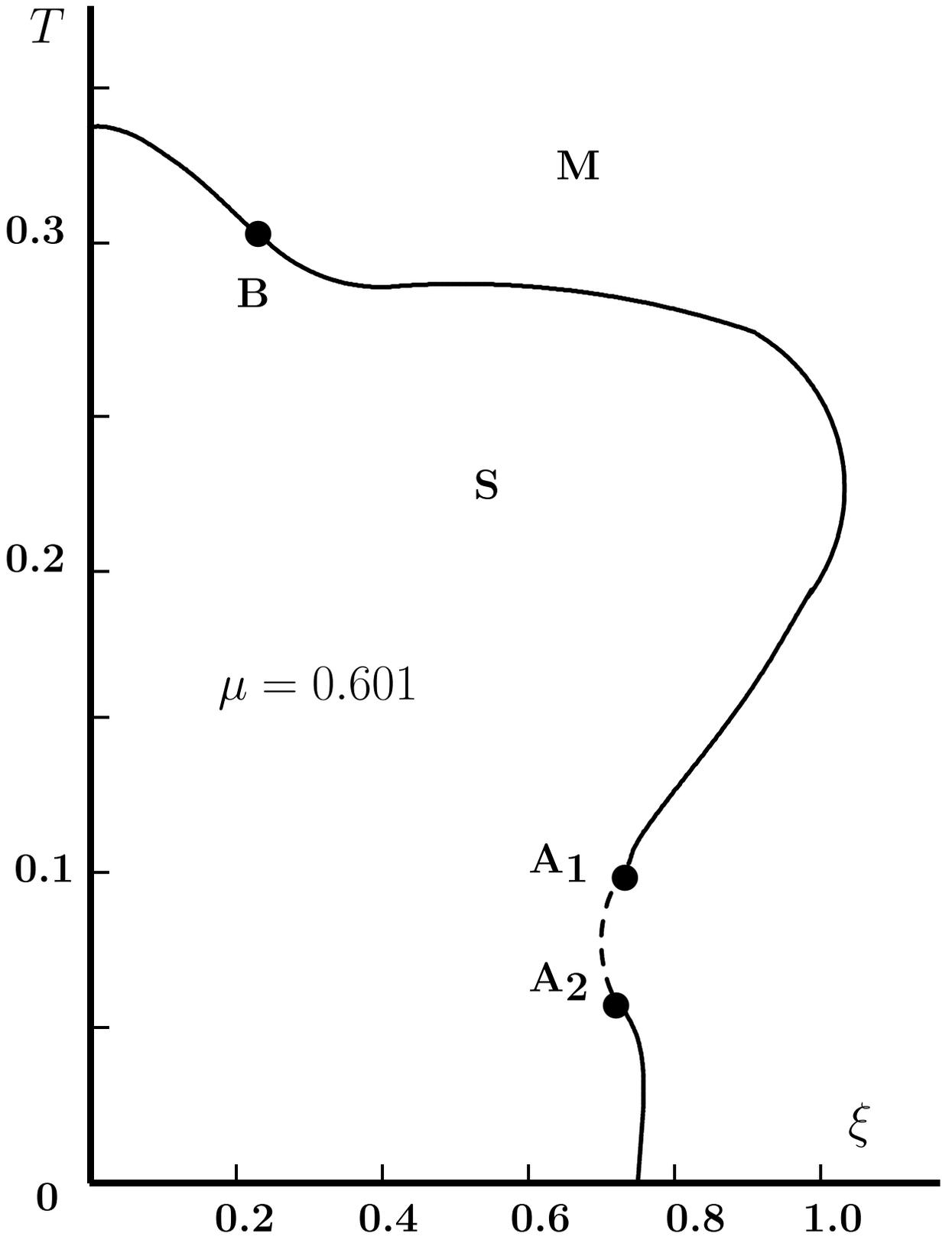
 \item[{\bf Fig.3}:] 
 The situation of $\mu=0.601$. ${\bf B}$ is a tricritical point. 
 The dashed line means 1st order transitions. Two tricritical points 
 ${\bf A_1}$ and ${\bf A_2}$ sit on the boundary between 1st and 2nd 
 order critical regions.
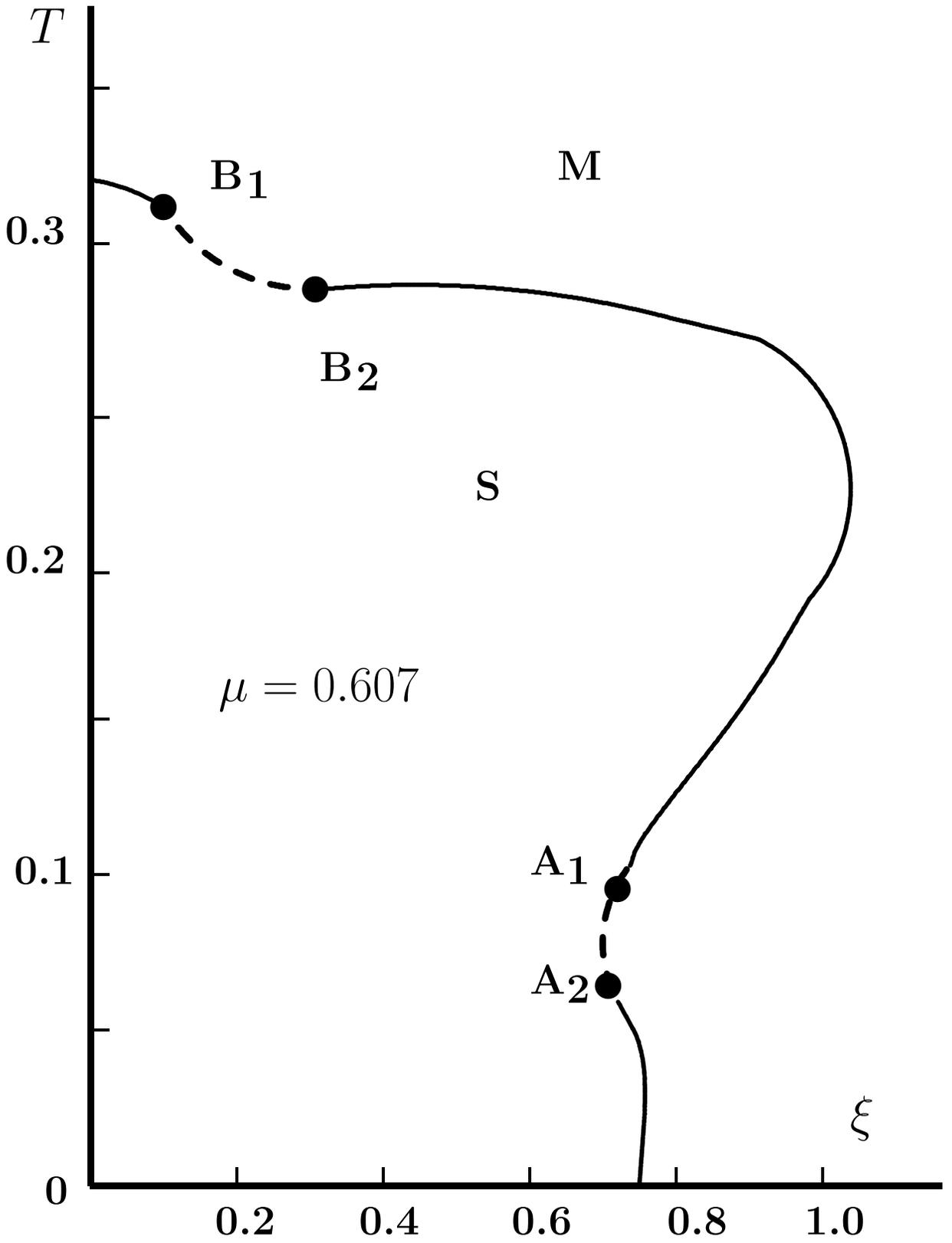
 \item[{\bf Fig.4}:]
 The phase diagram at $\mu=0.607$. ${\bf B_1},{\bf B_2},{\bf A_1},{\bf A_2}$ 
 are tricritical points.
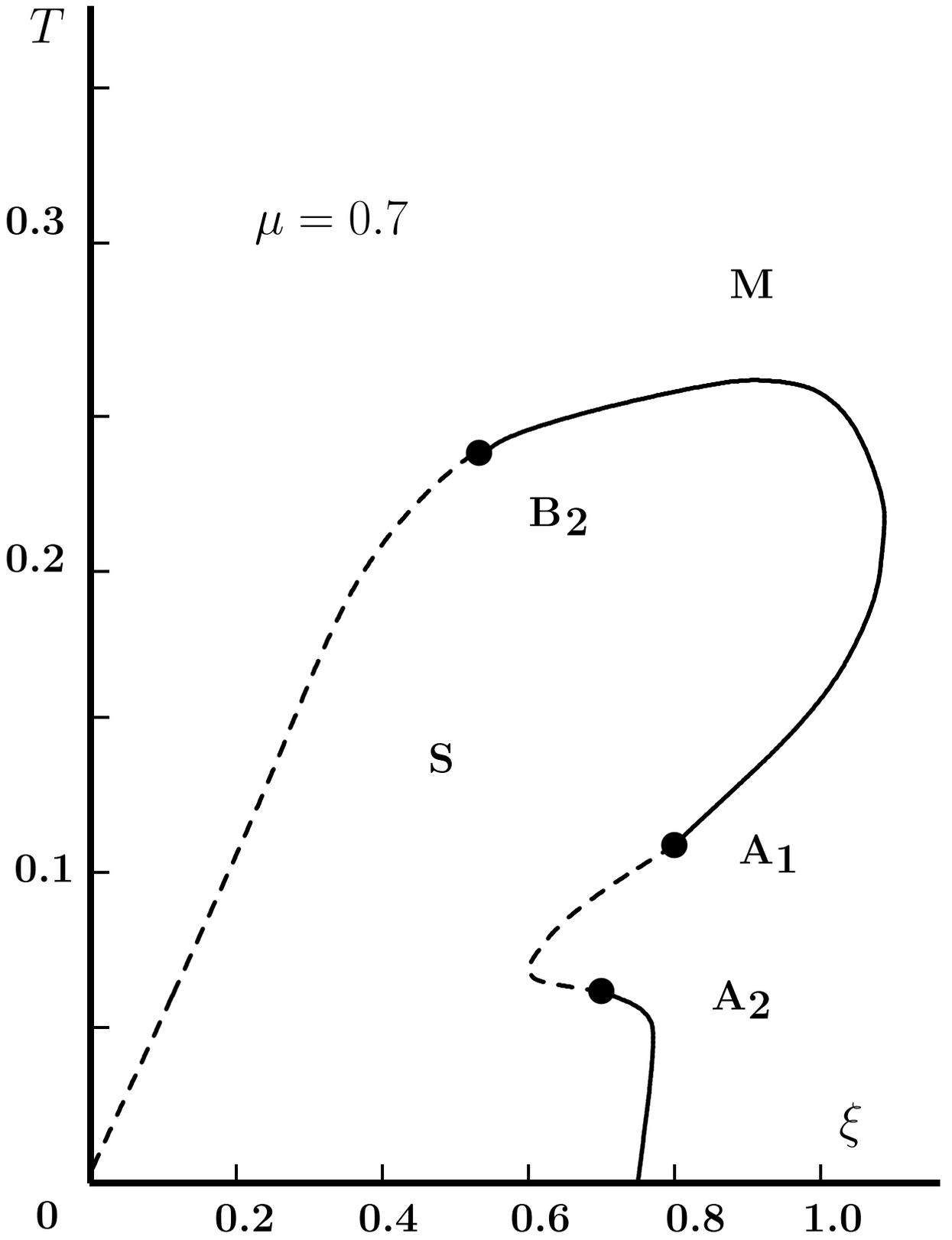
 \item[{\bf Fig.5}:]
 The phase diagram at $\mu=0.7$. ${\bf B_2},{\bf A_1},{\bf A_2}$ are 
 tricritical points.
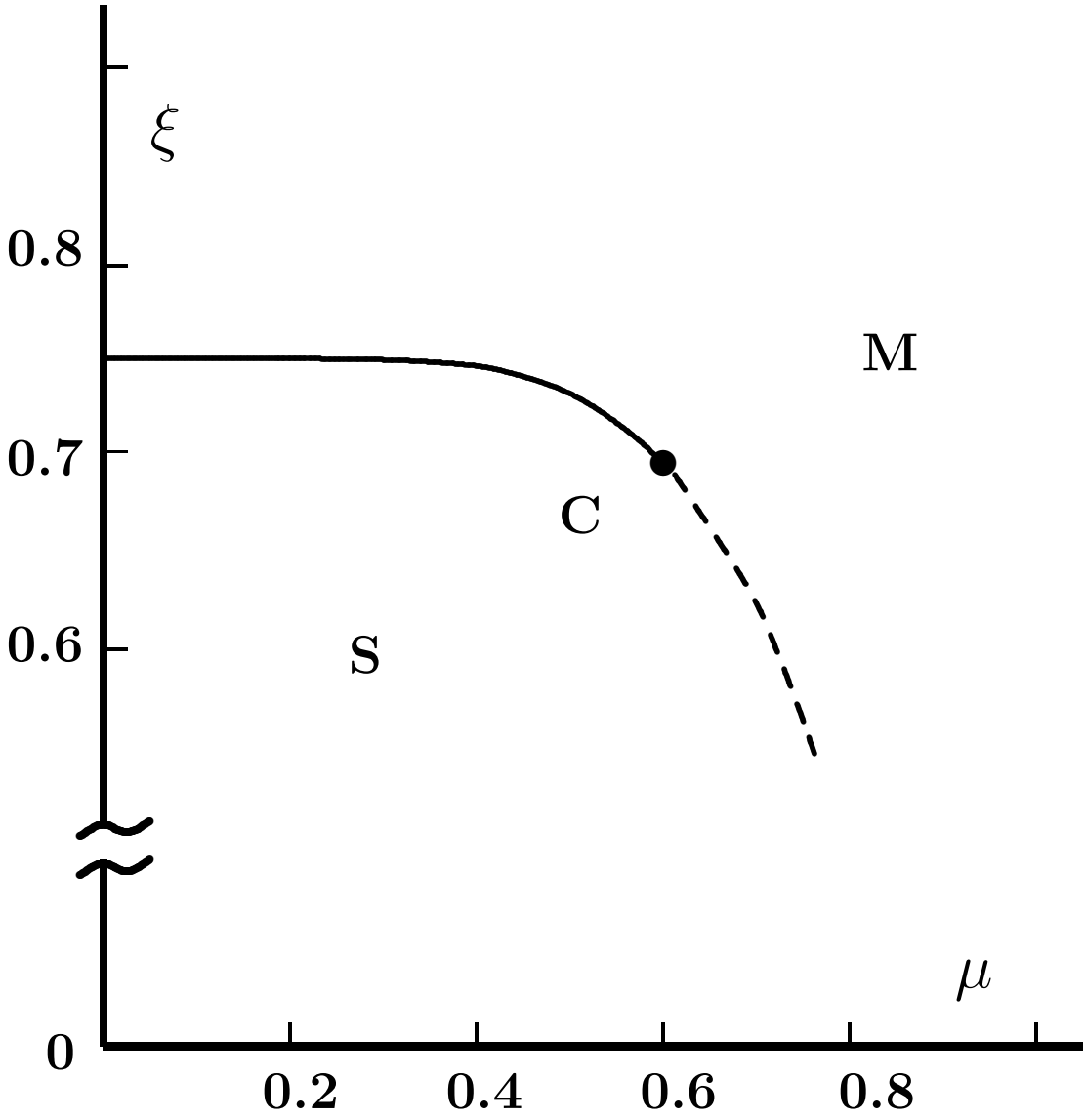
 \item[{\bf Fig.6}:]
 The phase line from weak $\mu$ expansion on $\mu$-$\xi$ plane. 
 ${\bf C}$ is tricritical. 
 \end{description}

\end{document}